\documentclass[12pt]{article}
\usepackage{epsfig}

\def\beq{\begin{equation}}
\def\eeq#1{\label{#1}\end{equation}}
\def\eeqn{\end{equation}}

\def\beqa{\begin{eqnarray}}
\def\eeqa#1{\label{#1}\end{eqnarray}}
\def\eeqan{\end{eqnarray}}

\let\bar=\overbar

\def\Dslash{\not{\hbox{\kern-4pt $D$}}}
\def\dslash{\not{\hbox{\kern-2pt $\del$}}}

\def\msb{{\bar{\ssstyle M \kern -1pt S}}}

\newcommand{\ve}[1]{\mbox{\boldmath $#1$}}
\newcommand \bea{\begin{eqnarray}}
\newcommand \eea{\end{eqnarray}}

\def\Title#1{\begin{center} {\Large {\bf #1} } \end{center}}

\begin{document}

\Title{Tensor Correlations and Pions in Dense Nuclear Matter}

\bigskip\bigskip


\begin{raggedright}  

{\it E. Olsson$^{1,2}$ and  C. J. Pethick$^2$\index{Pethick, C. J.} \\
\bigskip
\footnote{Uppsala Astronomical Observatory, Box 515, SE-75120 Uppsala, SWEDEN} 
Uppsala Astronomical Observatory, Sweden\\
\footnote{NORDITA, Blegdamsvej 17, DK-2100 Copenhagen, DENMARK}NORDITA, 
Denmark}
\bigskip\bigskip
\end{raggedright}

\section{Introduction}

Despite concerted efforts by many groups over a period of more than
three decades, the problem of stellar collapse, supernova production, and
neutron star formation is still not understood. The basic physical inputs
required for calculations of stellar collapse are, apart from an initial
stellar model, the equation of state of hot dense matter, and the rates of
neutrino production, absorption, and scattering processes. There are
tantalizing hints that, if the microscopic physical input differed from
what is presently used in simulations, the ``delayed" mechanism \cite{bethewilson},
in which the outgoing shock wave is revived by deposition of energy by
neutrinos streaming out of the stellar core, could account for core collapse
supernovae and the formation of neutron stars in the collapse of a massive 
star.  

One such hint comes from a relatively old calculation in which it was 
found that the allowance for pion-like excitations in the equation of state 
led to an increase in the explosion energy \cite{wilson1}. This indicates 
the importance of performing improved calculations of the equation of state 
in which pionic degrees of freedom are treated realistically.
A second hint comes from parametric studies of core-collapse models, which
indicate that the outcome of stellar collapse is sensitive to modest changes
(a factor of two, say) in the neutrino opacity \cite{janka}. The effects of 
some parts of the nucleon-nucleon interactions on rates of neutrino
processes in stellar collapse have been investigated by a number of groups, but
so far the tensor part of the interaction
has received scant attention, except in the work of Raffelt and colleagues
\cite{raffelt1}. Given what is known about the role of the tensor force in
finite nuclei, there is reason to believe that its effect on neutrino
processes in stellar collapse could be significant.  

Numerical simulations of stellar collapse provide motivation for 
taking up the question of how important tensor correlations are in nuclear 
and neutron matter. In addition, theoretical investigations of the equation of 
state of cold dense matter point to there being strong 
pionic correlations at densities only slightly above that of saturated nuclear
matter \cite{akmal}. In the latter work it was found that a state resembling a
neutral pion condensate was energetically favorable compared with one that
did not, contrary to the view accepted for the past quarter of a century that
pion condensation is ruled out at such densities by the strong 
central part of the nucleon-nucleon interaction in the spin-isospin channel.
The pionic correlations give rise to maxima in the static longitudinal 
structure factor for the nuclear
spin in neutron matter, and for the nucleon spin-isospin in nuclear
matter, as shown in Fig.\ \ref{fig:sl}. The definition of the structure factor
is given in Eq.\ (\ref{structure}) below.
 \begin{figure}[htb] \begin{center}
\epsfig{file=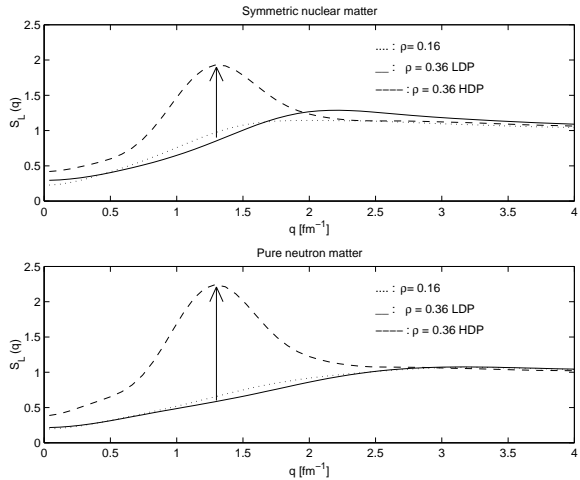,height=2.5 in} \caption{Static structure factor for the
longitudinal spin operator in neutron matter and for the longitudinal
spin-isospin operator in symmetric nuclear matter as  a function of wave
number, $q$. (After Ref.\ \cite{akmal}).} \label{fig:sl} \end{center}
\end{figure}
All these pieces of evidence point to the importance of investigating anew 
tensor correlations in dense matter. 

The aspect we shall focus on most here is that of neutrino processes, and how 
they are affected by non-central forces, of which the pion-exchange 
interaction between nucleons is an example. Our aim is to provide a general 
framework within which to describe neutrino processes, including
production, scattering and absorption. We shall make connections to Landau's 
theory of normal Fermi liquids, and indicate how the usual formulation, which 
assumes that the interaction is invariant under rotation of the spins of the 
particles, needs to be extended.

\section{Neutrino processes}

Rates of neutrino processes in dense matter are affected significantly by 
nucleon-nucleon interactions. Among early works on this subject we mention 
Refs. \cite{sawyer} and \cite{iwamoto} on neutrino scattering processes in 
degenerate matter. 
Rates of neutrino processes in dense matter were calculated with allowance 
for the effect of nucleon-nucleon interactions on the energies of the 
quasiparticles in Ref.\ \cite{prakashreddy}. In addition, the influence of 
the finite lifetime of excitations in the nuclear medium on rates of neutrino 
processes has been considered in Refs.\ \cite{raffelt1} and \cite{raffelt}. 
In the latter calculations, the non-central character of the nucleon-nucleon 
interaction plays an important role.  
Finally, in Ref.\ \cite{burrows}, the calculations of Refs.\ \cite{sawyer} and
\cite{iwamoto} have been extended to allow for partial degeneracy of the
nucleons, and for non-zero momentum transfers. In the latter work the 
screening of the weak interaction matrix elements by the central part of the 
nucleon-nucleon interaction was taken into account. 
Tensor correlations play an important role in determining the rates of 
neutrino processes in 
dense matter, both for the neutrino processes of importance in stellar 
collapse, as we shall explain in greater detail below, as well as for the 
modified Urca process which is important 
for the cooling of neutron stars. 

Since the weak coupling constant is small, it is generally an excellent
approximation to calculate the rate of neutrino processes in dense matter
using the standard golden-rule result for the transition rate. Furthermore,
the effects of other interactions, such as electromagnetic ones, between 
leptons and hadrons are generally small, and the 
expression for the rate $Q$ of transitions may be written in the form
\cite{raffelt}
\begin{equation}
Q \sim {G_F}^2 n \int
\frac{d^3\mathbf{q}d\omega}
{(2\pi)^4}S_{\mu\nu}(\mathbf{q}, \omega)N^{\mu\nu}(\mathbf{q},
\omega)
\end{equation}
where $G_F$ is the weak coupling constant, $n$ is the baryon density,
$S_{\mu\nu}(\mathbf{q}, \omega)$ is a dynamical structure factor for the
hadron system, and  $N^{\mu\nu}(\mathbf{q},
\omega)$ is a function that depends only on the leptons.  
The quantity $\bf q$ denotes the momentum transfer to the nucleon system,
and $\omega$ the energy transfer. The weak
interaction has the form of a product of current operators, either the vector
current or the axial vector one, for the leptonic and hadronic systems. For
definiteness, let us assume that the hadron system consists of 
non-relativistic nucleons. In this case, the leading contribution to a
current is the time component which, for the vector current, is proportional
to the nucleon density  and, for the axial vector
current, to the spin density. More
generally one must include relativistic effects, such as weak magnetism
(see e.g. Ref.\ \cite{horowitz}), and other hadronic degrees of freedom.

\subsection{Neutral-current processes}

To illustrate the qualitative effects of a non-central interaction, such as the
tensor interaction, on neutrino processes, we consider as an example scattering
of neutrinos by the weak neutral-current interaction. The structure factor of
interest in calculating the rate of the process is then the one for the spin
density. The tensor interaction has three qualitatively
different effects. These are conveniently examined for processes in which the
momentum and energy transfers are small compared with the typical energies and
momenta of a nucleon in the medium. One effect is to change the expectation
value of the spin of an excitation in the nuclear medium to a value different
from that for an isolated nucleon. This 
effect is well-known in the context of nuclear magnetic moments, where the
spin-orbit and tensor components of the nucleon-nucleon interaction modify the
nuclear moment
\cite{arima}. A second effect is to introduce into the effective interaction
between nucleon-like excitations a non-central component -- this modifies
screening phenomena. A third effect is that the spin operator, which, if one
neglects contributions due to exchange currents, is a one-body operator when
expressed in terms of nucleon degrees of freedom, acquires two- and higher-body
contributions for the excitations in the nuclear medium.    

\subsection{Landau theory}

To illustrate the effects of non-central interactions described above, we
consider the special case of long wavelengths, and temperatures low compared
with the Fermi temperature. In this regime the concepts and machinery of
Landau's theory of normal Fermi liquids may be applied \cite{landau}. We shall
examine the form of the spin response for a one-component system. 
As we mentioned above, this is
relevant for calculating the rates of neutral-current processes that proceed
via the axial vector contribution of the interaction.
If interactions are central, the static magnetic susceptibility of an
interacting Fermi liquid with a single fermion component may be expressed in
the form 
\begin{equation}
\frac{\chi}{\chi_0}=\frac{m^*/m}{1+G_0}
\end{equation}
where $\chi_0$ is the susceptibility of a non-interacting Fermi gas of the
same density, $m^*$ is the effective mass of a quasiparticle, and $G_0$ is the
Landau parameter describing the isotropic part of the spin-dependent component
 of the interaction. When there are non-central forces, this expression
must be modified, and we shall consider the three effects mentioned above in
turn.  

 The quasiparticle energy $\epsilon_{{\bf p}\ve{\scriptstyle\sigma}}$ is
defined quite generally by the equation
\begin{equation}
\epsilon_{{\bf p}  \ve{\scriptstyle\sigma}}=\frac{\delta E[n_{{\bf p}  
\ve{\scriptstyle\sigma}}]
}{\delta n_{{\bf p}  \ve{\scriptstyle\sigma}}}, 
\label{epsilon}
\end{equation}
where $E$ is the total energy and $n_{{\bf p}\ve{\scriptstyle\sigma}}$ is 
the quasiparticle distribution function.
The magnetic moment $\mu_{{\bf p} \ve{\scriptstyle\sigma}}$ of a 
quasiparticle is given by $({\mu}_ {\bf p  \ve{\scriptstyle\sigma}})_i = 
\partial \epsilon_{{\bf p} \ve{\scriptstyle\sigma}}/\partial B_i$.
From the requirement that the energy of a quasiparticle be
invariant under simultaneous rotations of the direction of the magnetic field
and of the quasiparticle momentum, it follows that the magnetic moment of a
quasiparticle at the Fermi surface must have the general form
\begin{equation}
\mu_i({\bf p})=\mu \sigma_i +(3/2)\mu_2 (\hat{b}_i \hat{b}_j -\delta_{ij}/3),
\end{equation}
where $\hat{\bf b}$ is a unit vector in the direction of the magnetic field,
and $\mu$ and $\mu_2$ are coefficients. 
The interaction between two quasiparticles $f_{{\bf p}
\ve{\scriptstyle\sigma},{\bf p'}  \ve{\scriptstyle\sigma}'}=\delta^2 
E[n_{{\bf p}  \ve{\scriptstyle\sigma}'}]/\delta n_{{\bf p}  
\ve{\scriptstyle\sigma}}\delta n_{{\bf p'}  \ve{\scriptstyle\sigma}'}
\label{f}$ may be written in the general form
\begin{equation}
f_{{\bf p} \ve{\scriptstyle\sigma},{\bf p'}  \ve{\scriptstyle\sigma}'}
=f_{{\bf p} {\bf p'}}  +
g_{{\bf p} {\bf p'}}{\ve{\sigma}}\cdot{\ve{\sigma}'}  
+  t_{{\bf p} {\bf p'}}(3{\ve{\sigma}}\cdot\hat{\bf 
k}\ {\ve{\sigma}'}\cdot\hat{\bf k}- 
{\ve{\sigma}}\cdot{\ve{\sigma}'}),
\end{equation}
where there is tensor contribution in addition to the usual expression
for a central interaction. 
If we assume that we may neglect all but the isotropic parts of the
quasiparticle interaction, one finds for the 
magnetic susceptibility the result \cite{olsson}
\begin{equation}
\frac{\chi}{\chi_0}=\frac{m^*}{m}\left(\frac
{\mu}{\mu_0}\right)^2\frac{1}{(1+G_0-K_0^2/8)}+ 
\frac{1}{5}\frac{m^*}{m}\left(\frac {\mu_2}{\mu_0}\right)^2 + 
\frac{\chi_{\rm M}}{\chi_0}, 
\end{equation}  
where $\mu_0$ is the magnetic moment of the particle in the absence of the 
medium, $K_0 = N(0)t$ is the tensor Landau parameter, $N(0)$
being the density of quasiparticle states at the Fermi surface, $G_0 = N(0)g$ 
and $\chi_M$ is the multipair contribution to the susceptibility.

One reason for the great success of Landau Fermi-liquid
theory for liquid $^3$He and for electrons in metals is that the forces
between particles are central to a good first approximation. In addition, the
quantities of greatest physical interest are the particle number density and
the spin density. What the result for the static susceptibility demonstrates 
is that many
more parameters are needed to characterize the long-wavelength properties of a
normal Fermi liquid with non-central forces, compared with the two needed for
a liquid with only central forces.

\subsection{Kinematics of neutrino processes}

Now let us consider scattering of neutrinos by dense matter. For the
simultaneous conservation of momentum and energy in the process, the
magnitude of the energy transfer $\omega$ to the neutrino must be less than
$cq$, where $\bf q$ is the momentum transfer. In other words, the
four-momentum transfer from the nuclear medium to the neutrino must be
space-like. For production of neutrino-antineutrino pairs via neutral-current
processes, the requirements are the opposite, since the total energy of the
pair must be greater than $cq$, $\bf q$ in this case being the total momentum
of the pair, or the four-momentum transfer must be time-like. It is
therefore of interest to explore which sorts of transitions in the nuclear
medium can contribute to the various classes of neutrino process. For
simplicity, let us again consider momentum transfers small compared with the
Fermi momentum of the nucleons, and temperatures low compared with the Fermi
temperature. The energy of a single quasiparticle-quasihole pair is given by 
$\omega =\epsilon_{\bf p} -\epsilon_{{\bf p}-{\bf q}} \approx \bf v_{\bf
p}\cdot{\bf q}$, where $\bf v_{\bf p} =\ve{\nabla}_{\bf p} \epsilon_{\bf p}$ 
always has a magnitude less than $v_{\rm F}q$,
where $v_{\rm F}$ is the velocity of a quasiparticle at the Fermi surface.
Thus single pair excitations in the nuclear medium always have
space-like four momentum. Consequently annihilation of a single
quasiparticle-quasihole pair cannot 
create a neutrino-antineutrino pair, while it  can contribute to scattering
of a neutrino. This clearly indicates the important role that excitations
containing two or more quasiparticle-quasihole states play in neutrino pair
emission. 

\subsection{Relative importance of single-pair and multi-pair states}
The discussion above indicates the different roles that the two sorts of
states play. One simple result may be obtained for operators ${\cal O}$ that
satisfy a local conservation law of the usual form 
\begin{equation}
\frac{\partial {\cal O}({\bf r})}{\partial t} +\ve{\nabla}\cdot{\bf j}
({\bf r}) =0,
\end{equation} 
where ${\bf j}({\bf r})$ is the associated current density. On taking 
the matrix element of this equation between an excited state $j$
and the ground state and Fourier transforming in space, one finds
\begin{equation}
\omega_{j0}({\cal O}_{\bf q})_{j0}= 
{\bf q}\cdot({\bf j}_{\bf q})_{j0}. 
\end{equation}
This shows that the matrix element of the density satisfies the condition
\begin{equation}
({\cal O}_{\bf q})_{j0}= 
\frac{{\bf q}\cdot({\bf j}_{\bf q})_{j0}}{\omega_{j0}}.
\label{matrixelement}
\end{equation}
Thus for an excited state with non-zero energy, the matrix element of the
operator tends to zero as $q$ tends to zero, provided only that the matrix
element of the current remains finite.
Therefore, since the vector current is conserved, there will be no multipair 
contributions to the time component of the current. However, the axial 
current is not conserved, and 
consequently multipair contributions will be present in general.

To determine how large these multipair 
contribution are, a number of approaches are possible. One is to 
calculate them directly from a nucleon-nucleon interaction and microscopic 
many-body theory. Another is to make use of sum rules to put bounds on 
multipair contributions. To illustrate the latter method, let us look at 
the structure function, defined by
\begin{equation}
S(q,\omega)=\sum_j |<j|{\cal{O}}_{\bf q}|0>|^2 \delta (\omega - \omega_{j0})=-{\rm Im } \chi(q,\omega)/\pi .
\end{equation}

In the limit $q \rightarrow 0$, only multipair states contribute to 
the static structure factor 
\begin{equation}
S(q)= \frac{1}{n}\sum_j |<j|{\cal{O}}_{\bf q}|0>|^2,
\label{structure}
\end{equation}
and to the energy-weighted sum
\begin{equation}
W(q)= \sum_j |<j|{\cal{O}}_{\bf q}|0>|^2\omega_{j0}.
\end{equation}
Another moment of interest is the static response function, $\chi(q,0)$, 
which is given by
\begin{equation}
\chi(q,0)= \sum_j 2|<j|{\cal{O}}_{\bf q}|0>|^2/\omega_{j0}.
\end{equation}

Let us denote the multipair contribution to the structure function by 
$S_{\rm M}(q,\omega)$. From the fact that the integrand is positive for 
positive $\omega$, it follows that    
\begin{equation}
\int_0^\infty \frac{S_{\rm M}(q,\omega)}{\omega}\left( \omega-\bar{\omega} 
\right)^2\geq 0,
\end{equation}
where the mean excitation energy $\bar{\omega} = W(q)/S(q)$.
This leads to the following condition on the multipair contribution to the 
static response function $\chi_{\rm M}$:
\begin{equation}
\chi_{\rm M}(0)\geq 2\ \frac{n S(q)}{\bar\omega}.
\end{equation}
If one were to take the $q \rightarrow 0$ limit of the calculations of the 
static spin structure factor and the  
energy-weighted sum calculated in Ref.\ \cite{akmal}, one would
be led to the conclusion that multipair excitations make up  
more than $\sim$60\% of the total static response function. This estimate is 
surprisingly large, but we should stress that
the calculations in Ref.\ \cite{akmal} were designed to shed light 
on correlations at non-zero wave numbers, where pionic correlations were 
found to enhance the static structure factor, and not to give accurate 
results for the long-wavelength 
response. It is possible that the calculations of Ref.\ \cite{akmal} 
overestimate the structure 
factor at small wave numbers. Consequently, improved calculations of the 
static structure function 
are needed before it will be possible to place better bounds on the 
magnitude of multipair contributions to the response.

\section{Concluding remarks}
From the discussion above it is apparent that tensor correlations 
potentially play an important role in the microscopic physics needed as  
input to simulations of stellar collapse and supernovae. Similar 
conclusions apply to other processes that we do not have space to consider 
here, among them charged current processes, such as the modified 
Urca process \cite{friman}, which is important for the cooling of 
neutron stars.

Many open problems remain. With respect to neutrino processes, improved 
expressions are needed 
for the modification of axial vector matrix elements in a nuclear 
medium, and for the Landau parameters \cite{brown}. To obtain an 
improved equation of state, the effects of pionic correlations need 
to be included in a realistic manner.  

\bigskip
We are grateful to A.\ Akmal for supplying us with tables of his 
results for the static structure factor and average excitation 
frequencies, and to 
S.\ Fantoni and V.\ R.\ Pandharipande for valuable discussions.     
Part of this work was carried out while the authors enjoyed the 
hospitality of the Institute of Nuclear Theory at the University
of Washington. 
E.O. thanks the school on Advanced Instrumentation and Measurements (AIM) and 
the EU Marie Curie Training Site at NORDITA for financial support.

\end{document}